\documentclass{article}

\usepackage[final]{neurips_2025_ml4ps}

\usepackage[utf8]{inputenc} % allow utf-8 input
\usepackage[T1]{fontenc}    % use 8-bit T1 fonts
\usepackage{hyperref}       % hyperlinks
\usepackage{url}            % simple URL typesetting
\usepackage{booktabs}       % professional-quality tables
\usepackage{amsfonts}       % blackboard math symbols
\usepackage{nicefrac}       % compact symbols for 1/2, etc.
\usepackage{microtype}      % microtypography
\usepackage{xcolor}         % colors
\usepackage[pdftex]{graphicx} 
\usepackage{textgreek}
\usepackage{orcidlink}

\usepackage{subcaption}

\newcommand{\tighttable}{%
  \small
  \setlength{\tabcolsep}{4pt}%
  \renewcommand{\arraystretch}{0.95}%
}

\title{Transformer Embeddings for Fast Microlensing Inference}

\author{%
    \textbf{Nolan Smyth}$^{1,2,3}$\orcidlink{0000-0002-8454-3015} \and 
    \textbf{Laurence Perreault-Levasseur}$^{1,2,3,4}$\orcidlink{0000-0003-3544-3939}
    \textbf{Yashar Hezaveh}$^{1,2,3,4}$\orcidlink{0000-0002-8669-5733} 
    \\
    $^{1}$Universit{\'e} de Montr{\'e}al \quad $^{2}$Mila \quad $^{3}$Ciela Institute \quad $^{4}$CCA, Flatiron Institute \\
    \texttt{\{nolan.smyth, laurence.perreault.levasseur, yashar.hezaveh\}@umontreal.ca}}

\begin{document}

\maketitle

\begin{abstract}
  The search for free-floating planets (FFPs) is a key science driver for upcoming microlensing surveys like the Nancy Grace Roman Galactic Exoplanet Survey. These rogue worlds are typically detected via short-duration microlensing events, the characterization of which often requires analyzing noisy, irregularly-sampled observations. We present a pipeline for this task using simulation-based inference. We use a Transformer encoder to learn a compressed summary representation of the raw time-series data, which in turn conditions a neural posterior estimator. We demonstrate that our method produces accurate and well-calibrated posteriors over three orders of magnitude faster than traditional methods. We also demonstrate its performance on KMT-BLG-2019-2073, a short-duration FFP candidate event.
\end{abstract}

\section{Introduction}
\label{sec:intro}

Free-floating planets (FFPs) may be the most ubiquitous type of terrestrial-mass exoplanet, potentially outnumbering their bound counterparts by a factor of more than 20 \cite{sumiFreeFloatingPlanetMass2023}. Low-mass FFPs primarily form in planetary disks and are subsequently ejected from their system of origin \cite{zwartOriginFreefloatingObjects2024, colemanPredictingGalacticPopulation2024}. Due to their negligible electromagnetic radiation and lack of host star, FFPs are extremely difficult to detect. Gravitational microlensing is the most promising technique to find these rogue worlds \cite{gaudiMicrolensingSurveysExoplanets2012}. Microlensing occurs when a compact foreground mass passes near the line of sight to a background star, warping the light around the lens. The multiple images produced at the observer are unresolved, resulting in a temporary smooth, achromatic magnification of the background star. 

The Nancy Grace Roman Space Telescope is expected to detect thousands of FFPs \cite{johnsonPredictionsNancyGrace2020}. This will usher in a new era of exoplanet demographics, illuminating the origins of these elusive objects \cite{gouldFreeFloatingPlanetsEinstein2022a,deroccoReconstructingFreefloatingPlanet2025}. Crucial to this effort is rapid characterization of microlensing signals. Traditional methods like Markov Chain Monte Carlo (MCMC) are computationally expensive and do not scale well to the billions of light curves Roman will deliver. While anomaly detection pipelines are expected to significantly filter this dataset, a significant proportion will still require detailed characterization. Simulation-Based Inference (SBI) provides a powerful framework for efficient posterior estimation by amortizing the cost of simulation. 

Neural Posterior Estimation (NPE) is a SBI approach where a neural network is trained to learn the Bayesian Posterior $p(\theta|x)$ over the model parameters $\theta$ given the observed data $x$, bypassing the need for likelihood evaluations. A key benefit of this amortized approach is that training is a one-time cost, after which inference for new observations is extremely fast, requiring only a single forward pass through the trained network. Previous work has used amortized SBI for binary microlensing events, motivated by the computational cost of the forward model \cite{zhangRealTimeLikelihoodFreeInference2021}. This work used a 1D ResNet with a Gated Recurrent Unit as an embedding network for fixed-length, regularly-sampled data. We found that a similar, well-calibrated recurrent network trained on regularly-sampled light-curves failed catastrophically when even minor data gaps were introduced. This is a classic example of the well-known difficulty of handling distributional shifts in time-series data \cite{gagnon-audetWOODSBenchmarksOutofDistribution2023, shuklaSurveyPrinciplesModels2021}.

\begin{figure}[t]
\centering
\includegraphics[width=0.9\textwidth]{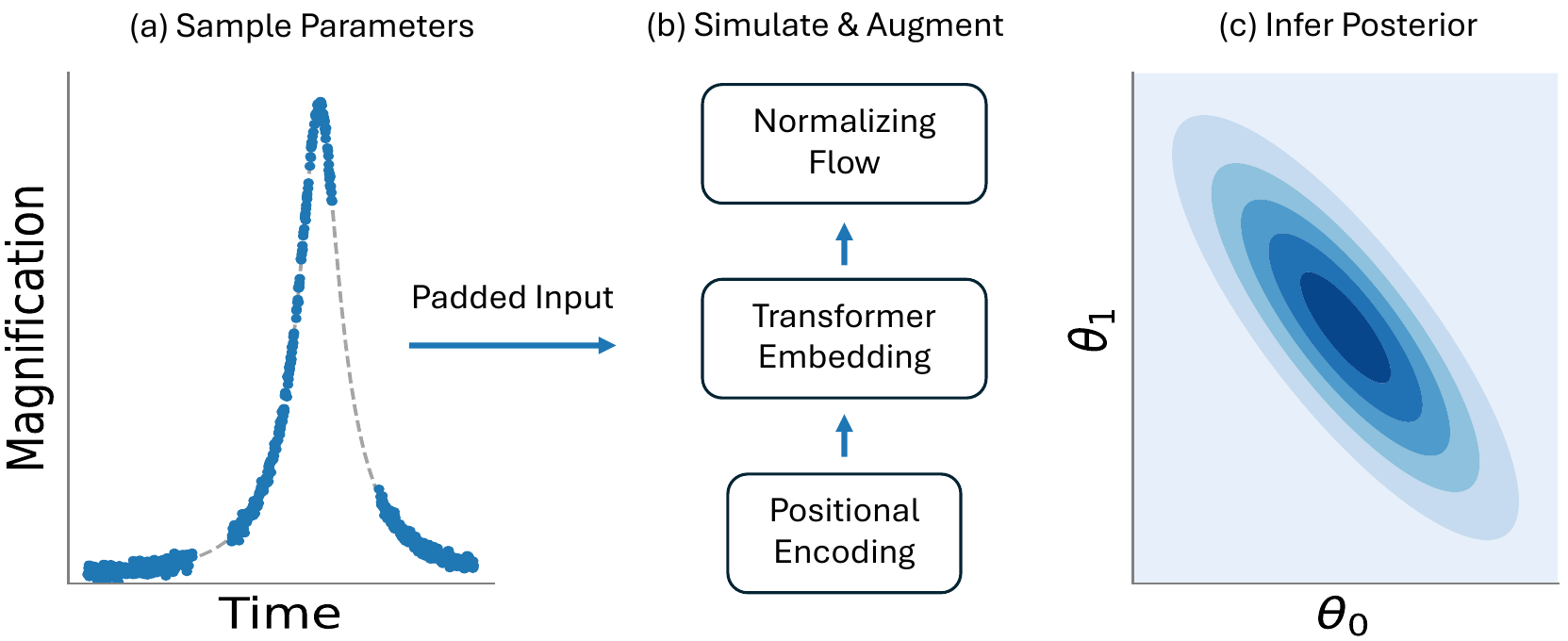}
  \caption{Our SBI pipeline. Parameters $\theta$ are drawn to simulate a light curve, which is then augmented with data gaps, dropout, and noise. The resulting observation $x$ is fed into a Transformer encoder to produce an embedding that conditions a normalizing flow approximating the posterior $p(\theta|x)$.}
  \label{fig:method}
  \vspace{-6mm}
\end{figure}

In this work, we develop an end-to-end SBI pipeline for microlensing parameter estimation that naturally handles variable length, sparse, noisy, and irregularly sampled time-series data.\footnote{All the code used in this work is available at: https://github.com/NolanSmyth/sbi\_microlensing\_transformers.} While previous work has applied transformer encoders for astrophysical time series data (e.g. \cite{koblischkeSpectraFMTuningStellar2024, zhangMavenMultimodalFoundation2024}), the primary contribution of this work is to demonstrate a robust and scalable solution to this specific challenge, enabling fast and accurate inference directly from raw, time-series data without the need for imputation or complex pre-processing. We use a Transformer encoder \cite{vaswaniAttentionAllYou2023}, as the self-attention mechanism is particularly well-suited to this type of data. A schematic of this pipeline is shown in Figure \ref{fig:method}. We validate our pipeline, showing it produces accurate and well-calibrated posteriors across a wide range of event morphologies. We also demonstrate the pipeline's performance on a real-world FFP candidate, KMT-2019-BLG-2073 \cite{kimKMT2019BLG2073FourthFreeFloatingPlanet2021}, recovering posterior distributions consistent with the publicly available data \cite{ kimKMT2019BLG2073FourthFreeFloatingPlanet2021, albrowDifferenceImagingPhotometry2009}.

\section{Method}
\label{sec:method}

We simulate microlensing events using a finite-source, point-lens (FSPL) model implemented with the \texttt{VBMicrolensing} package \cite{bozzaVBBinaryLensingPublicPackage2018a}. The parameters, collectively denoted by $\theta$, are: the time of closest approach $t_0$; the minimum impact parameter in units of the Einstein radius $u_0$; the Einstein crossing time $t_E$; the normalized source radius $\rho \equiv \theta_\star /\theta_E$; and the source flux fraction $f_s \equiv \frac{F_{\rm{source}}}{F_{\rm{source}} + F_{\rm{blend}}}$. The total observed flux at time $t$, normalized to the baseline, is $F(t) = f_s A(t,\theta) + (1-f_s)$, where $A$ is the magnification. All timescales are in units of days. 

To train a robust network, we employ on-the-fly data augmentation. For each training sample $\theta_i$ drawn from the prior shown in Table \ref{tab:priors}, we generate a dense, noiseless light-curve. We then apply a sequence of random augmentations: \textbf{Seasonal Gaps:} Introduce 0 to 3 gaps, each with a length of 1 to 10 days. \textbf{Random Dropout:} Remove a random fraction (0\%-60\%) of the remaining points to simulate data quality cuts. \textbf{Noise Injection:} Add Gaussian photometric noise, with $\sigma$ for each light-curve drawn uniformly from $[0.001, 0.02]$ in relative flux units. 

Each light-curve is represented as a padded sequence of length $L=1000$. Each timestep has three channels: $x_i = (t_i^{\rm{norm}}, F_i, \sigma_i), i = 1, ..., L.$ Times are normalized to $[-1,1]$ over a window of $T=20$ days and $\sigma_i$ is the per-point photometric uncertainty. Padded timesteps are masked with a value of $-2$. To ensure the network trains on meaningful signals, we filter these augmented light-curves using a set of recoverability criteria: at least 5 data points must lie within $t_E/2$ of the peak, $t_0$; at least 5 points must lie more than $2t_E$ from the peak to establish a baseline magnitude; the peak magnification must be at least $5\times$ larger than the mean per-point flux error, $F(t_0)/\sigma > 5$. In a full analysis pipeline, there would be a process for identifying the most promising light-curves - these criteria are meant to serve as a quality assurance check to avoid the network being trained or evaluated on light-curves that would not have a recoverable microlensing signal. 

Our NPE pipeline uses a standard Transformer encoder \cite{vaswaniAttentionAllYou2023} to map the input sequence $x \in \mathbb{R}^{L\times3}$ to a summary vector $z \in \mathbb{R}^d$. The network consists of an input projection layer, sinusoidal positional encoding, and a stack of 6 Transformer layers with 8 attention heads, a model dimension of 256, and a feed-forward dimension of 512. To aggregate the variable-length sequence output, we perform normalized average pooling over the unmasked timesteps. We then use a Masked Autoregressive Flow, as implemented by the \texttt{sbi} package \cite{greenbergAutomaticPosteriorTransformation2019, boeltsSbiReloadedToolkit2025}, to model the posterior distribution $p(\theta|x)$, conditioned on the Transformer embedding. The network is trained by minimizing the negative log-likelihood of the posterior on pairs of simulations $\{ (\theta_i, x_i) \}$. We trained our network on 80,000 simulated events, plus an additional 20,000 for validation during training. We use the Adam optimizer \cite{kingmaAdamMethodStochastic2017} with an initial learning rate of $10^{-4}$ and a \texttt{ReduceLROnPlateau} scheduler that reduces the learning rate by a factor of 0.5 with a patience of 10 epochs. All training and evaluation was conducted in $\sim 20$ hours on a single Nvidia H100 GPU using 16GB of memory.

\section{Results}
\label{sec:results}

\begin{figure}[t]
  \centering

  \includegraphics[width=0.9\linewidth]{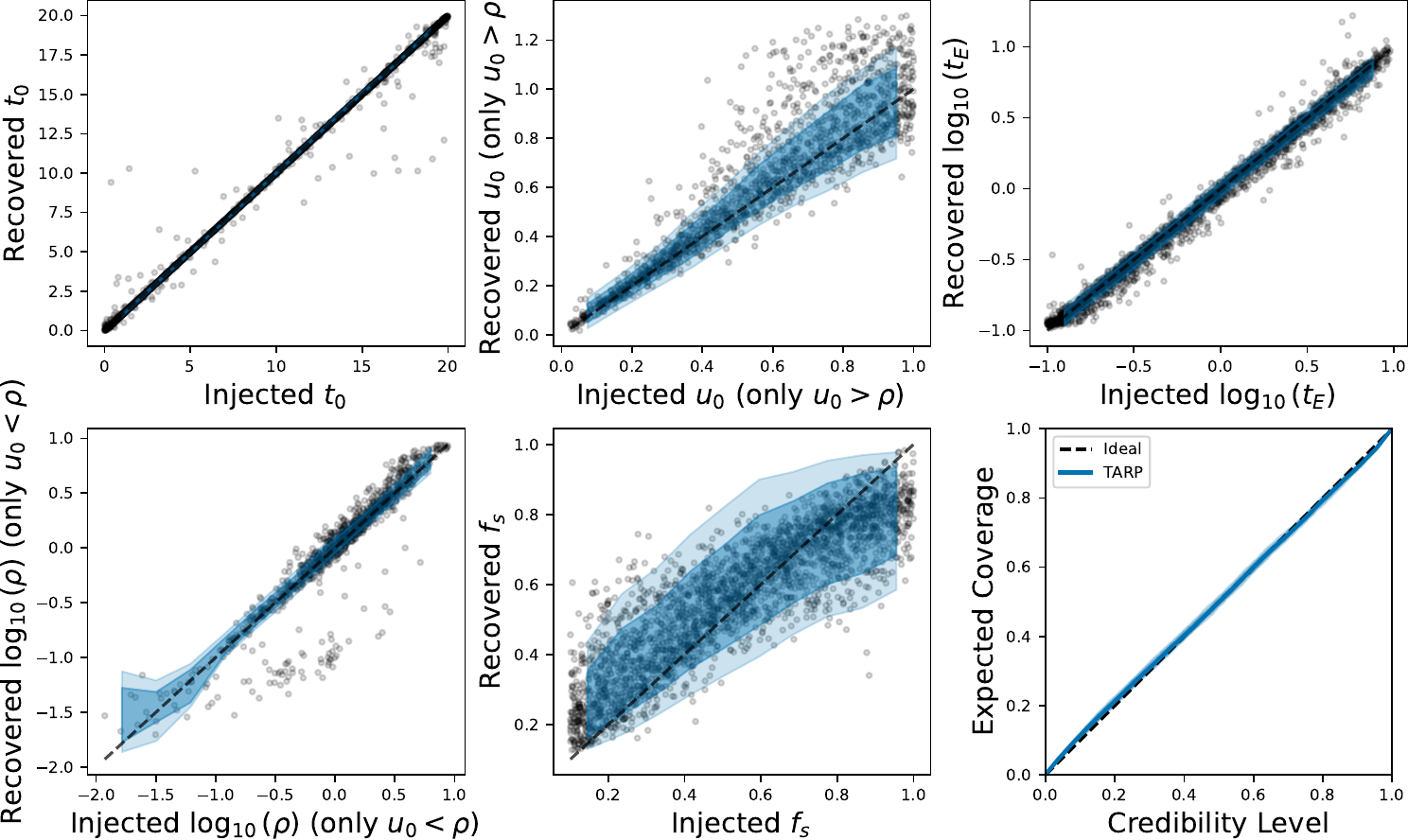}
  \caption{Injected–recovered plots show recovered median posterior parameters. The shaded regions show the 16-84 and 5-95 percentile ranges, averaged per bin. The TARP diagnostic is shown in the bottom right, demonstrating excellent calibration.}
  \label{fig:calibration}
\vspace{-5mm}
  
\end{figure}

We assess the calibration of our trained network on a test set of 5,000 simulated events, drawing 5,000 posterior samples from each event. Figure \ref{fig:calibration} shows the injected-vs-recovered values for each parameter and the overall model calibration. Each panel, except the bottom right, shows the posterior median recovered values against the true injected values. For the impact parameter $u_0$ and source size $\rho$, we plot only the regimes where they are identifiable. For point-source-like events $(u_0 > \rho)$, $u_0$ is well-recovered, while for events with strong finite-source effects ($u_0 < \rho)$, $\rho$ is well-recovered. The bottom right panel shows a TARP (Tests of Accuracy with Random Points) diagnostic \cite{lemosSamplingBasedAccuracyTesting2023}. Calibrated TARP is both a necessary and sufficient condition for an accurate posterior estimator. Crucially, the NPE yields a more than $10^4$ factor speedup compared to ensemble sampling as implemented in \texttt{emcee}\cite{foreman-mackeyEmceeMCMCHammer2013} to generate the same number of posterior samples (see Appendix \ref{app:A} for more details). 

\begin{figure}[t]
  \centering
  \includegraphics[width=0.78\linewidth]{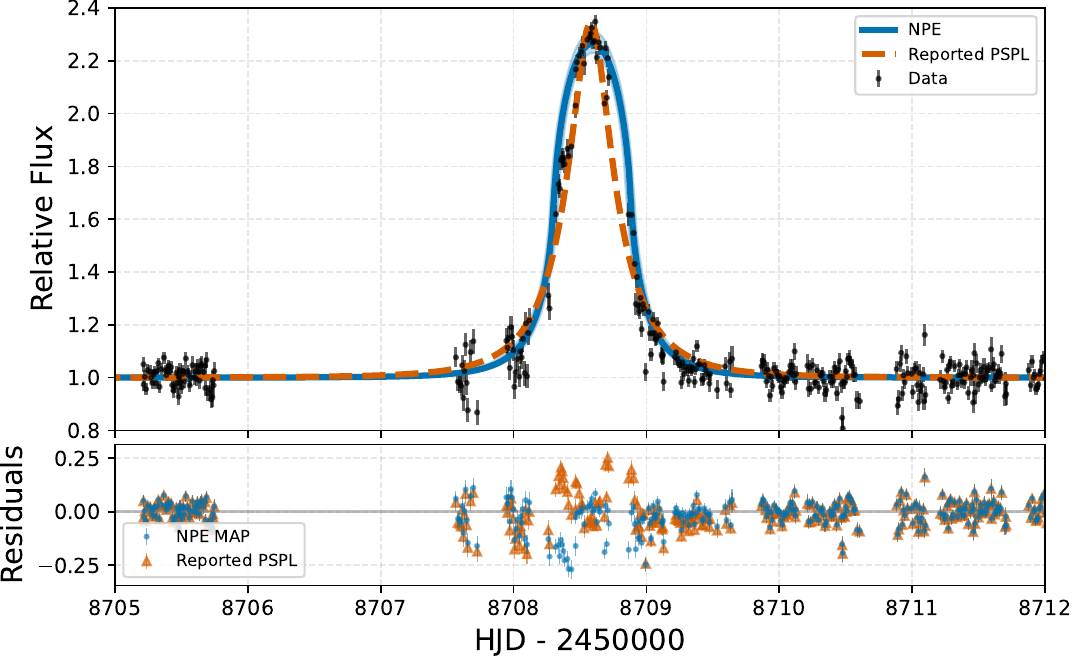}
  \caption{Data and model fits to KMT-2019-BLG-2073 using pySIS (I-band). Time is shown as Heliocentric Julian Date (HJD). The shaded band shows the 5-95 percentile range recovered by the NPE, while the dashed line is the reported PSPL fit.}
  \label{fig:kmt}
  \vspace{-5mm}
\end{figure}

We also applied the pipeline to KMT-2019-BLG-2073 \cite{kimKMT2019BLG2073FourthFreeFloatingPlanet2021}, a short-duration microlensing event with pronounced finite-source effects, classified as a "likely FFP candidate". We used publicly available I-band data from the KMTNet survey's pySIS pipeline \cite{albrowDifferenceImagingPhotometry2009}. \footnote{https://kmtnet.kasi.re.kr/ulens/event/2019/view.php?event=KMT-2019-BLG-2073.} We convert the raw magnitudes to an absolute flux scale with a reported zero point of $18.15$,  $F_{\rm abs}=10^{-0.4 (I-18.15)}$, and then normalize the light-curve by the baseline flux. Since the fit reported in \cite{kimKMT2019BLG2073FourthFreeFloatingPlanet2021} was performed on data from KMTNet's TLC pipeline, it does not provide a direct comparison for the publicly available pySIS data we analyze here. Such differences between different photometric extraction pipelines is important and should be considered depending on the application. The recovered posteriors are shown in Table \ref{tab:kmtpost}. Figure \ref{fig:kmt} shows the data, light-curve recovered by the NPE, and the best-fit point-source-point-lens (PSPL) model. The FSPL model provides an excellent fit around the peak where finite-source effects dominate, with residuals comparable to or smaller than those of the PSPL fit.

\begin{table}[t]
  \centering
  \caption{Priors and recovered posteriors.}
  \vspace{-2mm}

  \begin{subtable}[t]{0.48\linewidth}
    \centering
    \caption{Priors}
    \label{tab:priors}
    {\tighttable
    \begin{tabular}{lc}
      \toprule
      Parameter & Prior \\
      \midrule
      $t_0$ & $\mathrm{Uniform}(0,\,20)$ \\
      $u_0$ & $\mathrm{Uniform}(0,\,1.5)$ \\
      $t_E$ & $\mathrm{LogUniform}(0.1,\,20)$ \\
      $\rho$ & $\mathrm{LogUniform}(10^{-2},\,10)$ \\
      $f_s$ & $\mathrm{Uniform}(0.1,\,1)$ \\
      \bottomrule
    \end{tabular}
    }
  \end{subtable}\hfill
  \begin{subtable}[t]{0.48\linewidth}
    \centering
    \caption{Recovered posteriors for KMT-2019-BLG-2073}
    \label{tab:kmtpost}
    {\tighttable
    \begin{tabular}{lcc}
      \toprule
      Parameter & Recovered value & Reported Fit \\
      \midrule
      $t_0$ & $8708.60^{+0.02}_{-0.02}$ & 8708.58\\
      $u_0$ & $0.20^{+0.11}_{-0.10}$ & 0.32\\
      $t_E$ & $0.355^{+0.034}_{-0.031}$ & 0.50\\
      $\rho$ & $0.832^{+0.080}_{-0.090}$ & N/A \\
      $f_s$ & $0.82^{+0.11}_{-0.13}$ & 0.61\\
      \bottomrule
    \end{tabular}
    }
  \end{subtable}
\end{table}

\section{Discussion and Limitations}
\label{sec:discussion}

Our SBI approach with a Transformer embedding demonstrates a powerful, automated method for microlensing characterization. In deployment, the priors should be informed by the expected distribution of detectable events, given by a hierarchical galactic and lens population model \cite{koshimotoParametricGalacticModel2021a}. This is beyond the scope of the current work, but important as SBI methods can be sensitive to distributional shifts \cite{filippRobustnessNeuralRatio2024}. 

Furthermore, the model is trained using Gaussian noise, which is a good approximation of the fundamental Poisson measurement uncertainty but does not capture more complex, systematic noise sources present in real detectors or insidious false-positives. False-positive signals are ubiquitous due to stellar variability, magnetic activity, and other transients that can mimic temporary increased flux in a source. This issue also arises with traditional methods (see e.g. \cite{kunimotoSearchingFreeFloatingPlanets2024b, mrozTESSFreefloatingPlanet2024, yangHowRareAre2024}). It will be crucial to also model these false positives and either include them in the detection pipeline, or model them for accurate inference. Additionally, the model is trained on a fixed 20-day window, tailored to short-duration events associated with low-mass FFPs, but limiting its applicability to long-duration events from more massive lenses like brown dwarfs or intermediate mass black holes \cite{perkinsHintsAnomalousLens2025,perkinsDisentanglingBlackHole2024,deroccoRogueWorldsMeet2024}. This is straightforward to address with an extended prior, depending on the use-case. Our application to KMT-2019-BLG-2073 serves as a promising proof-of-concept. However, a more extensive validation on a larger, diverse set of real-world events is required to fully establish the method's reliability and is a key direction for future work.

There are several avenues for future architectural exploration. Ablation studies would help determine the optimal model size and embedding dimension. We utilized sinusoidal positional encoding added to the input. While the timestamps of the inputs should contain all the necessary positional information, we noted that this encoding resulted in faster convergence during training. Quantifying this improvement and comparing encoding schemes could potentially offer significant benefits. Also, while our use of normalized average pooling over unmasked tokens is simple and effective, other pooling or aggregation mechanisms may better capture information from the features of the microlensing peak.

Lastly, we note that our pipeline is complimentary to automated anomaly detection pipelines (see e.g. \cite{vianaLensNetEnhancingRealtime2025}) that are intended to reduce the number of light-curves to be manually analyzed.

\section{Conclusion}
\label{sec:conclusion}
We have presented a robust and calibrated SBI pipeline for microlensing analysis. By leveraging a Transformer encoder, our method can directly process sparse, noisy, and irregularly-sampled time-series data, overcoming a major limitation of amortized inference approaches in this domain. This work provides a powerful, scalable tool for analyzing the large datasets. This will be crucial for upcoming surveys like the Roman Space Telescope, helping to accelerate the discovery and characterization of free-floating planets.

\bibliographystyle{unsrt} 
\bibliography{references} 

\newpage

\appendix
\section{Benchmarks}
\label{app:A} 

\begin{figure}[t]
  \centering
  \begin{subfigure}[t]{0.49\linewidth}
    \centering
    \includegraphics[width=\linewidth]{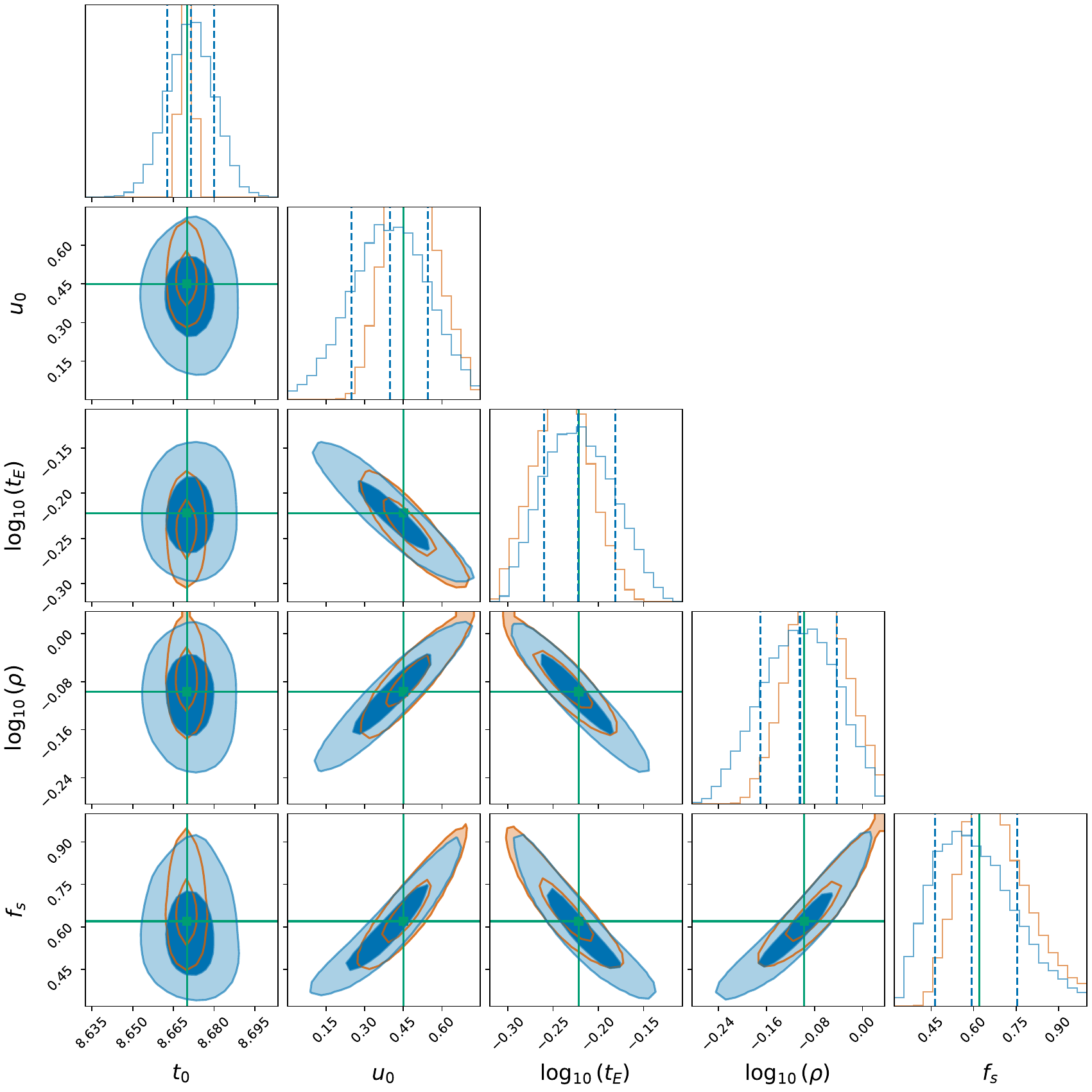}
    \caption{Posterior comparison. MCMC results are displayed in orange and NPE results in blue. The true parameter values are shown in green.}
    \label{fig:corner_overlay}
  \end{subfigure}\hfill
  \begin{subfigure}[t]{0.49\linewidth}
    \centering
    \includegraphics[width=\linewidth]{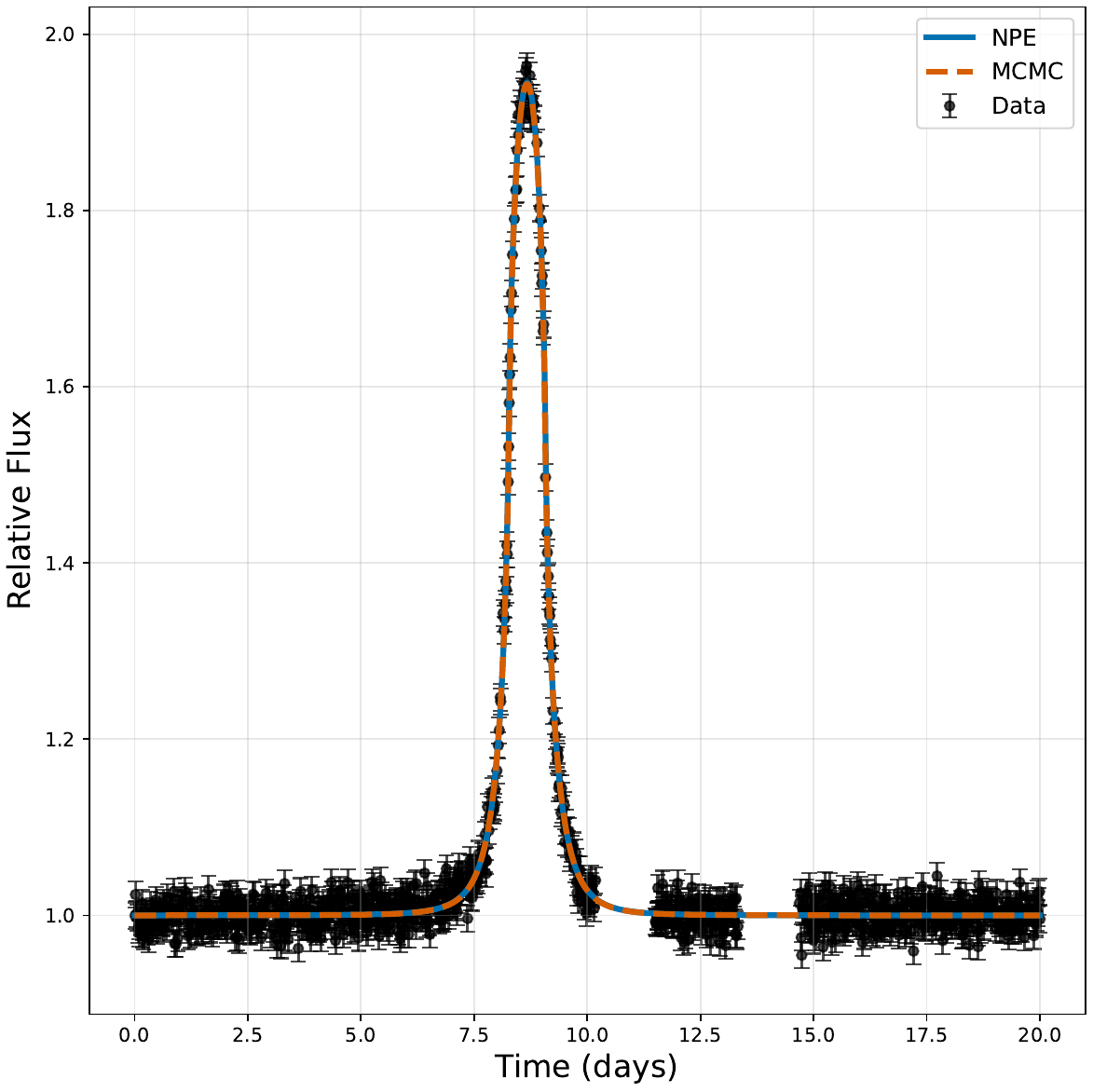}
    \caption{Simulated light-curve and recovered models for the MCMC and NPE. The 16-84 percentile confidence intervals are shown, but are practically negligible as both models fit the data extremely well.}
    \label{fig:lc_compare}
  \end{subfigure}
  \caption{NPE and MCMC posteriors on the same simulated event.}
  \label{fig:npe_mcmc_compare}
\end{figure}

We show the results of the NPE and a traditional MCMC for a simulated lightcurve in Figure \ref{fig:npe_mcmc_compare}. The MCMC uses 32 walkers, 5,000 burn-in steps, followed by 10,000 samples. We directly generate 15,000 samples from the NPE. Drawing 15,000 samples from the trained NPE takes just 0.08 seconds on a GPU and 0.82 seconds on a CPU. For comparison, running the MCMC sampler for 15,000 steps took 959 seconds on a CPU, resulting in an inference speedup factor of about $1.2 \times 10^3$ for a single light curve on the same hardware and a factor of more than $10^4$ when GPU accelerated.

As shown in Figure \ref{fig:corner_overlay}, the NPE posteriors are in excellent agreement with the MCMC results, with a slight broadening visible in some parameters. This is consequence of the amortized inference. The network has learned a approximation to ensure robust calibration across the entire range of possible light curves, but is well-calibrated across the entire prior space, as revealed by TARP diagnostic. The posterior estimate for a single event could quickly be refined using importance sampling or as seeding to a down-stream MCMC if necessary. The light-curves recovered by both the NPE and MCMC fit the data extremely well, as shown in Figure \ref{fig:lc_compare}. The extremely narrow bands corresponding to the 16-84 percentile light-curves also demonstrates the inherent degeneracies of microlensing recovered by the model.

% \begin{ack}
% Use unnumbered first level headings for the acknowledgments. All acknowledgments
% go at the end of the paper before the list of references.
% \end{ack}

%%%%%%%%%%%%%%%%%%%%%%%%%%%%%%%%%%%%%%%%%%%%%%%%%%%%%%%%%%%%

\end{document}